\documentstyle[12pt,aasms4]{article}

\def\ltsim{\, {}^<_\sim \,}
\def\etal{{\it et al.}}

\def\eg{{\it e.g.}}

\def\mag{{$\,$mag}}
\def\deg{\ifmmode^\circ\else$^\circ$\fi}    

\def\hper{\ifmmode \rlap.^{h}\else $\rlap{.}^h$\fi} 
\def\sper{\ifmmode \rlap.^{s}\else $\rlap{.}^s$\fi}    

\def\Min{${}^{\prime}$\llap{.}}

\def\deg{${}^\circ$}

\def\gtsim{ \,{}^>_\sim\, }
\def\bmv{\hbox{\it B--V\/}}

\def\ngc#1{NGC$\,$#1}

\def\today{\number\year\space \ifcase\month\or
  January\or February\or March\or April\or May\or June\or
  July\or August\or September\or October\or November\or December\fi
  \space\number\day}
\def\now{\number\year\space \ifcase\month\or
  January\or February\or March\or April\or May\or June\or
  July\or August\or September\or October\or November\or December\fi
  \space\number\day .\number\time}
\reversemarginpar

\lefthead{Stetson}
\righthead{Photometric Standard Stars}

\begin{document}

\title{Homogeneous Photometry for Star Clusters and Resolved Galaxies. II.
Photometric Standard Stars}

\author{Peter B. Stetson\footnote{Visiting Astronomer, Kitt Peak National
Observatory and Cerro Tololo Inter-American Observatory, which are
operated by the Association of Universities for Research in Astronomy,
Inc.\ (AURA) under cooperative agreement with the National Science
Foundation.}$^,$\footnote{ Visiting Astronomer, Canada-France-Hawaii Telescope
operated by the National Research Council of Canada, the Centre National de
la Recherche Scientifique de France and the University of
Hawaii.}$^,$\footnote{Guest User, Canadian Astronomy Data Centre, which is
operated by the Herzberg Institute of Astrophysics, National Research
Council of Canada.}$^,$\footnote{Guest investigator of the UK Astronomy Data
Centre.}}

\affil{Dominion Astrophysical Observatory, Herzberg Institute of
Astrophysics, National Research Council, 5071 West Saanich Road, Victoria,
British Columbia V8X 4M6, Canada \\Electronic mail:
Peter.Stetson@nrc.ca}

\vspace{2 in}
\begin{abstract}

Stars appearing in CCD images obtained over 224 nights during the course
of 69 observing runs have been calibrated to the Johnson/Kron-Cousins {\it
BVRI\/} photometric system defined by the equatorial standards of Landolt
(1992, AJ, 104, 340).  More than 15,000 stars suitable for use as
photometric standards have been identified, where ``suitable'' means that
the star has been observed five or more times during photometric
conditions and has a standard error of the mean magnitude less than
0.02\mag\ in at least two of the four bandpasses, and shows no significant
evidence of intrinsic variability.  Many of these stars are in the same
fields as Landolt's equatorial standards or Graham's (1982, \pasp, 94,
244) southern E-region standards, but are considerably fainter.  This
enhances the value of those fields for the calibration of photometry
obtained with large telescopes.  Other standards have been defined in
fields containing popular objects of astrophysical interest, such as star
clusters and famous galaxies, extending Landolt-system calibrators to
declinations far from the equator and to stars of sub-Solar chemical
abundances.  I intend to continue to improve and enlarge this set of
photometric standard stars as more observing runs are reduced.  The full
current database of photometric indices is being made freely available via
a site on the World-Wide Web, or by direct request to the author.
Although the contents of the database will evolve in detail, at any given
time it should represent the largest sample of precise {\it BVRI\/}
broad-band photometric standards available anywhere.

\end{abstract}

\keywords{Standards; Stars:  general}

\section{Introduction}

Accurate photometry with modern detectors on large telescopes is hampered
by the scarcity of suitable photometric standard stars.  At present, the
largest and most definitive collection of fundamental standard stars in
the Johnson {\it UBV\/} Kron-Cousins {\it RI\/} broad-band photometric
system is that of Landolt (1992), which consists of 526 stars that are
mostly quite close to the celestial equator.  However, if one restricts
oneself only to those stars that were observed a minimum of five times
each (for instance), with standard errors of less than 0.02\mag\ in both
$V$ and \bmv\ (say), then the total number of Landolt's ``good" standards
is reduced to 318.  Of these, perhaps of order 200 are appropriate for use
with a 2.5-m telescope ($V \gtsim 12$); maybe $\sim\,$130 can be used with
a 4-m telescope ($V \gtsim 13)$; and $\ltsim\,$40 are suitable for use
with an 8-m telescope or a 2.4-m telescope in space ($V \gtsim 14.5$).
Graham (1982) has published a list of some 103 stars with {\it UBVRI\/}
photometry in nine fields at declination --45\deg; if one again considers
only those stars having at least five observations and standard errors in
$V$ and \bmv\ less than 0.02\mag, the number of ``good'' Graham standards
is reduced to some 61, of which only 11 are fainter than $V = 12$.

There are places on the sky where several standards can be imaged onto a
CCD at the same time, but many of the Landolt and Graham stars are
comparatively isolated, so that trying to observe a diverse sample of
standards over a range of colors and airmasses with a CCD can be quite
inefficient.   Furthermore, since these standards are primarily equatorial
or far south, they never reach the zenith at many good terrestrial
observing sites, and cannot cover the same range of azimuth as many
scientifically interesting targets.   Observers trying to make the most of
their large-telescope time are often reluctant to undertake large slews
from the science target to one or more standard fields more than a 
few times per night.  Another drawback of Landolt's and Graham's standards
is that few or no Population~II stars are included.

However, when one does observe fundamental standards like Landolt's or
Graham's with a CCD, one usually gets for free the images of nearby stars,
most of which are fainter than the official standards.  I expect that most
CCD photometrists have toyed with the notion of combining these
serendipitous observations of neighbor stars for the purpose of defining
new, fainter standards, and this is what I have begun to do.  As of this
date (Spring 2000) I have combined photometric data from a total of 69
observing runs consisting of 224 individual nights, of which 135 nights
were completely clear, while on the remaining 89 nights observations were
obtained through thin cloud during at least part of the night.  (CCD
observations made through cloud can contribute to the precision of
photometric indices provided that each image contains either fundamental
standards or secondary standards that have also been observed under
photometric conditions on numerous occasions.  Differential photometry
relative to the brighter, well-established stars reduces the random errors
of the mean magnitudes estimated for the fainter stars in the same
field.)  These observations have been made by many different observers
using ten telescopes at five sites (Kitt Peak National Observatory:  4-m,
2.1-m, 0.9-m; Cerro Tololo Inter-American Observatory: 4-m, 1.5-m, 0.9-m;
La Palma:  Isaac Newton Telescope, Jacobus Kapteyn Telescope;
Canada-France-Hawaii Telescope; Wyoming Infrared Observatory) over the
period 1983--1999.  Many of these observations were made by me or my
collaborators, but I have also obtained data for many of these observing
runs through the excellent services of the Isaac Newton Group Archive and
the Canadian Astronomy Data Centre.

In addition to photometric measurements for faint neighbors of Landolt and
Graham standards, I have defined new standard sequences on the same
photometric system in fields where the presence of an astrophysically
interesting object (\eg, a star cluster or a nearby galaxy) has led to the
field's being observed several times during the observing runs at my
disposal.  In the case of star clusters or dwarf galaxies very near the
Milky Way, many of new standards will actually be members of the science
target.  In the case of more distant galaxies and other types of
extragalactic object, the new standard
stars obviously belong to the Galactic foreground.  At the present
moment, the available data permit the definition of more than 15,000 primary
and secondary standards in 198 fields, where the following criteria are
satisfied:  at least five independent observations under photometric
conditions {\it and\/} standard errors of the mean magnitude smaller than
0.02\mag\ in {\it at least two\/} of {\it BVRI\/}, and no evidence for
intrinsic variation in excess of 0.05\mag, root-mean-square, based upon
consideration of all available bandpasses.  The Johnson $U$ bandpass is
not much observed with CCDs due to a variety of inconveniences, such as
the low and highly wavelength-dependent relative quantum efficiency of
many CCDs at these short wavelengths.  Although I do have and have
tabulated some $U$-band observations for a number of these stars, I have
not considered the availability of $U$ data to be relevant in making the
decision whether a given star warrants being considered a photometric
standard for my present purposes.

Lists of these standards are available to interested photometrists via the
World-Wide Web or by direct communication with me.  The available data
are:  digital finding charts (FITS-format images) on a common
half-arcsecond-per-pixel scale, with $x$ increasing east and $y$
increasing north; ASCII files with astrometric positions---both absolute
right ascensions and declinations, and relative (x,y) positions in the
finding charts; and lists of photometry, consisting of mean apparent
magnitudes in {\it UBVRI\/}, the standard errors of those quantities, the
number of independent observations in each filter (the number of
observations made on photometric occasions and the total number of
observations, including those made through thin cloud, are both
tabulated), and a measure of the intrinsic root-mean-square photometric
variation.  All of these observations have been placed on the system of
Landolt (1992) with an accuracy of order 0.001\mag\ in the mean.  It is my
intention to keep the database up to date as additional observing runs
become reduced, so the random photometric errors should go down and the
number of individual standards and independent fields may be expected to
grow with time.  However, at any given moment the instantaneous state of
the database should represent the largest and most precise sample of {\it
BVRI\/} broad-band photometric standards available anywhere.

\section{Detailed Discussion}

At the moment, the total set of CCD observations considered here consists
of some 1,092,401 individual magnitude measurements for 28,552 stars.  The
instrumental magnitudes are based entirely on synthetic aperture
photometry (bright, isolated stars) or profile-fitting photometry with
aperture growth-curve corrections (fainter stars, or those with neighbors
less than a few arcseconds away) obtained with CCDs and extracted by means
of software written by me (Stetson 1987, 1990, 1993, 1994).  The
instrumental magnitudes are transformed to the standard system using
nightly equations that generally include linear and quadratic color terms
as well as linear extinction terms.  Whenever practical, mean color
coefficients are determined for all the nights of a given observing run
with a particular instrumental setup.  However, extinction coefficients
and photometric zero points are determined on a night-by-night basis,
except for a few cases where the range of airmass spanned by the
observations is too small for a meaningful extinction measurement; in such
cases mean extinction coefficients for the site are imposed.  The
equations for non-photometric nights do not model the effects of
extinction.  Instead a separate photometric zero point for each frame is
determined from measurements of at least two standard stars included
within that frame; color terms determined from photometric nights during
the same run and/or from individual frames containing standards that span
a broad range of color are employed just as for the photometric nights.
After the transformation equations for all nights have been determined,
all the observations for each star are collected and transformed to the
best possible magnitudes in $U$, $B$, $V$, $R$, and $I$ based on a
simultaneous least-squares optimization involving all available data for
the star.

The whole process is iterated.  Initially transformation equations are
determined only from observations of the fundamental standard stars.  Then
standard-system magnitudes can be derived for other stars contained within
the same fields as the fundamental standards, and for stars in other
program fields that were observed on photometric nights.  The subset of
these stars that meet the criteria mentioned above, {\it viz.\/} at least
five observations made under photometric conditions {\it and\/} standard
errors smaller than 0.02\mag\ in at least two of the four {\it BVRI\/}
filters, and no significant evidence of intrinsic variability, may now be
considered to be additional standards.  Improved transformations are then
re-determined using this enlarged set of standard stars.  Starting with
this second iteration, the newly defined ``standards'' allow the inclusion
of non-photometric observations for the former program fields, increasing
the precision (but not the accuracy) of their derived photometric
magnitudes.  Another iteration of this process is undertaken every time a
new observing run is added to the database, resulting in some new standard
stars, more precise mean magnitudes for the previously existing standard
stars, and occasionally the loss of a putative standard if the new
observations suggest intrinsic variability.

The fundamental basis for the photometric system employed here is that of
Landolt (1992) consisting of (mainly) equatorial standards observed in
{\it UBVRI\/} with photomultipliers.  I have augmented this primary set of
reference stars with the data in Landolt (1973; photomultiplier-based {\it
UBV\/} observations that are apparently independent of those of Landolt
1992, unlike the observations in Landolt 1983, which appear to be a subset
of those included in the 1992 catalog); Landolt (1983; a very few stars
that were not republished in the 1992 paper); Graham (1982;
photomultiplier {\it UBVRI\/} photometry of stars in the E regions at
declination --45\deg) and Graham (1981; photomultiplier {\it UBVRI\/}
photometry of a standard sequence near the spiral galaxy \ngc{300});
W.~E.~Harris (unpublished; photomultiplier {\it UBV\/} photometry of stars
in the equatorial open cluster M11 = \ngc{6705}); and L.~Davis
(unpublished; CCD {\it UBVRI\/} photometry of stars in the Kitt Peak
consortium fields in the star clusters \ngc{4147}, 2419, 6341 = M92, 7006,
and 7790; Christian \etal\ 1985).

All of these data must be assumed {\it a priori\/} to be on effectively the
same photometric system as Landolt (1992) --- within the errors --- with
two exceptions.  (1)~There are enough stars in common between Landolt
(1973) and Landolt (1992) that a direct comparison of the two systems can
be undertaken in $U$, $B$, and $V$.  In fact, I base this comparison on
{\it only\/} those stars that are common among Landolt (1973), Landolt
(1992), {\it and\/} the set of Landolt stars included among my
observations.  This restriction is made just in case any difference
between Landolt (1973) and Landolt (1992) might depend in some systematic
way on the stars' magnitudes, colors, right ascensions, or other
properties; if such should be the case, obviously we want to know the
value for any (1992) {\it minus\/} (1973) difference that would be
appropriate specifically for the sorts of stars considered here.  When the
comparison is made, I find that the Landolt (1992) magnitudes differ from
those of Landolt (1973) by --0.0034$\pm$0.0011\mag\ (standard error of the
mean difference) in $V$, --0.0026$\pm$0.0013\mag\ in $B$, and
+0.0022$\pm$0.0023\mag\ in $U$, based upon 81 stars common to all three
data sets.  Landolt's 1973 {\it UBV\/} magnitudes have been adjusted by these
offsets and combined with his 1992 data.  (2) According to Davis (private
communication) her data for \ngc{7790} were taken under dubious
photometric conditions. The way in which these data are included will be
described below.  

The assumption that the remaining Graham, Harris, and Davis data are on
essentially the same system as Landolt (1992), at least within the
standard errors of the available data sets, can be tested {\it a
posteriori}, as I will now describe.  Specifically, after each iteration I
compare my photometry with Landolt's for those stars where (a)~Landolt has
at least four observations and a standard error of the mean magnitude less
than 0.03\mag\ in a given filter, {\it and\/} (b)~I have at least four
observations and a standard error of the mean magnitude less than
0.03\mag\ in the same filter, {\it and\/} (c)~the star shows no evidence
in my data for intrinsic variability greater than 0.05\mag,
root-mean-square, in all filters considered together.  (Selection criteria
more restrictive than these resulted in a sample size too small to be very
meaningful.) Any net difference remaining between my weighted average
results and the combined results of Landolt (1992) and (1973) for stars
meeting these criteria is evaluated and added to all my magnitudes,
forcing my photometric system to be identically equal, in the mean, to
that of Landolt with a high level of accuracy.  After the most recent
iteration these corrections were all less than 0.0005\mag\ in $B$, $V$,
$R$, and $I$, with standard errors of the correction better than
0.0013\mag\ in each case, based on 144 stars in $B$, 144 stars in $V$, 30
stars in $R$, and 79 stars in $I$; in $U$ the correction  was
$0.0009\,\pm\,0.0084$\mag\ based on only 3 stars.  Figures 1--4 show the
differences between my photometry and Landolt's for these stars versus
magnitude and color.  The observed root-mean-square magnitude residuals
between Landolt's results and mine exceed the quadrature sum of both our
estimated standard errors by less than 10\%.  This leaves very little room
for systematic errors due to neglected high-order transformation terms
occasioned by, for instance, filter-bandpass mismatch.  

In fact, to the naked eye, some seemingly systematic differences between
my photometry and Landolt's may be seen in Figs.~1--4.  For instance, in
Fig.~1 it seems that for $10\,\ltsim\,B\,\ltsim\,11.5$ my $B$-band
magnitudes are fainter than Landolt's, while for
$11.5\,\ltsim\,B\,\ltsim\,12.0$ my $B$ magnitudes are brighter.  Similarly,
my $B$ magnitudes for the bluest stars ($\bmv < 0.00$) may be slightly
fainter, on average, than Landolt's.  If such behavior is real, it would
imply a subtle systematic nonlinearity either Landolt's photometry or
mine.  In either case, the nonlinearity would have to be a {\it
collective\/} property of many devices, since Landolt used a number of
different photomultipliers and cold boxes, while my results have certainly
been based on a large number of different CCD and filter combinations.  In
each case, all data for each detector were placed on a common photometric
system using appropriate transformation models.  It is noteworthy that the
apparently systematic differences in the $B$-band photometry are {\it
not\/} duplicated in either the $V$ band or the $I$ band, while the plots
for each of these other bandpasses have idiosyncracies likewise not
reflected in the other filters.  I cannot come up with a plausible
physical mechanism that would produce this variety of effects
systematically across an ensemble of detectors of either technology.  In the
absence of more definitive data, it seems most likely that these seeming
deviations are the result of small-number statistics and the propensity of
the human eye for finding patterns even in random data.

After I have thus forced my mean results onto Landolt's system, the net
differences between Davis's unpublished magnitudes for \ngc{7790} stars
and my results for the same stars are determined and applied to her mean
magnitudes; these corrections, which are of order a few hundredths of a
magnitude, place Davis's \ngc{7790} data on the same system as mine, in
the mean, which is---via the previous step---the same as Landolt's.
Finally, a weighted average of my derived photometric magnitudes and the
previous ones is determined for all stars in common.  To the extent that
the Graham results, the Harris results, and the rest of Davis's results
may {\it not\/} be inherently on the Landolt system, these weighted mean
magnitudes will not be on exactly the Landolt system either.  However,
they will be much {\it closer\/} to the Landolt system inasmuch as my
observations generally greatly outnumber the previous ones.  In fact,
these other data sets turn out to be fairly close to the Landolt system in
comparison to their standard errors, as Table~1 shows.  Here I have
tabulated the robust mean magnitude differences and standard errors of the
mean differences  for {\it all\/} stars common to my and the previous data
sets, without regard to the number of observations, the standard errors,
or any evidence of variablility.  Only two elements of the table reveal
systematic differences as large as 0.01\mag, and it is to be expected that
the overall set of my observations combined with the previous ones will
differ from the mean Landolt system by amounts much less than this.  The
line of mean differences for the comparison to Landolt's photometry
exemplifies the ultimate uncertainty of placing my photometry on his
system:  it represents the distinction between (a)~comparing only those
stars that he and I both measured ``well,'' in some sense, which has zero
net difference after the procedure of the previous paragraph, and
(b)~comparing {\it all\/} stars common to the two data sets, which yields
the differences in Table~1.

\section{General Discussion}

For purposes of the remaining discussion, I will regard a star as being
suitable to serve as a photometric standard if, when all my observations
have been combined with all the data from the Landolt, Graham, Harris, and
Davis star lists, it has been observed at least five times and has a
standard error less than 0.02\mag\ in {\it at least two\/} of $B$, $V$,
$R$, and $I$, {\it and\/}, when a weighted average is taken of the
standard deviations of the measured magnitudes in all available filters,
the implied net intrinsic variation is less than 0.05\mag.  At the
moment, 15,419 stars in 198 fields satisfy these criteria.  Among these,
96 fields have at least five standards in at least two of the filters, and
21 have at least five standards in all four filters within the area of a
single CCD field.

Table~2 is a very partial listing of some of the fields containing
standard stars defined in this way, intended only to give some sense of
the declinations, field sizes, numbers of standards, and types of contexts
that are available.  The table lists the equatorial coordinates of each
field for equinox 2000, the rectangular dimensions spanned by the standard
stars in the field in units of arcminutes of right ascension and
declination, and the number of stars with standard-quality magnitudes,
defined by the criteria given above, in each of the four principal
bandpasses.  Observations were obtained in all four of the {\it BVRI\/}
filters during only a very few of the 69 observing runs treated here.  The
reader will therefore notice that generally there are not equal numbers of
standards in all four filters.  This is an unavoidable result of the fact
that different fields were observed during different runs employing CCDs
of different projected angular size and different combinations of two or
more filters.  It is also true that, although the absence of close, bright
neighbors was one of the selection criteria for potential new standards,
some of these stars may be too crowded for use with telescopes of short
focal length or under conditions of particularly poor seeing.  Similarly,
some of these stars will be too bright for the largest telescopes or too
faint for the smallest ones.  Nevertheless, With reasonable care,
interested photometrists should be able to find in the database a good
selection of suitable standard fields as they plan observations utilizing
any particular equipment and combination of bandpasses, for any given range
of right ascension and north or south declinations.

Since the precision of the photometry of the Landolt, Graham, and Kitt
Peak consortium standards has now been improved by the addition of many
more observations, but more especially because numerous new standards have
been added in many of these same fields, astronomers who want to can now
make retroactive improvements to the photometric accuracy of any studies
they have already undertaken that used the previously published
standards.  In addition, the many new standards that have been defined
with apparent magnitudes as much as 6\mag\ fainter than those previously
available offer a new opportunity for accurate future photometry with the
largest telescopes.  The much larger number of fields over a wide range of
declinations greatly simplifies the task of finding standard fields
relatively near specific science targets and allows for improved extinction
determinations, including the possibility of testing for extinction
variations as a function of azimuth.  Finally, the provision of standard
sequences on a common system within the very fields of some of the most
popular science targets offers a new level of homogeneity in the
intercomparison of stellar populations---a principal goal of the present
series of papers.  (Paper I is Stetson, Hesser \& Smecker-Hane 1998).

One of the more noteworthy aspects of this work is that, with the
inclusion of a number of globular clusters among the standard fields, for
the first time we have now available a single, homogeneous system of
broad-band photometry based on standard stars spanning Populations I and
II.  To be strictly rigorous, it is not correct to state that Landolt's
(1992) photometric system has now been extended to Population II.  In
order to claim that, I would have to be able to say that we now know
accurately what magnitudes Arlo Landolt would have measured for any given
random star with {\it his\/} photomultipliers and filters during the
period 1977--1991.  This is something I cannot claim.  The most that I can
say is that I have defined a system based on a somewhat more democratic
principal:  these are the magnitudes that an arbitrary astronomer using
typical and commonly used CCD/filter combinations would be most likely to
obtain for a large, heterogenous sample of stars spanning a broad range of
metal abundance and evolutionary state, after doing his or her best to
transform the observed magnitudes in a consistent way to the system of
Landolt (1992).  These data represent a new photometric system which spans
Populations~I and II, but which very closely equates to the Landolt
system, in the mean, at the Population~I end.

As stated in the Introduction, finding charts, astrometric positions, and
photometric indices may be obtained from a World-Wide Web site hosted by
the Canadian Astronomy Data
Centre\footnote{http://cadcwww.hia.nrc.ca/standards; under the heading
``Photometric Standards'' click on ``\underline{Stetson}''.}, or by direct
request to the author.  At the present moment, 53 of the 198 fields are
completely documented and ready for use by the general astronomical
public.  In general, these are the fields that have the most standards in
the most filters.  However, all 198 fields are listed in the complete
version of Table~2 that is available at the Web site; a complete list of
potential standard fields will also be provided by the author on request.
If a particular photometrist has a need for one of the standard fields
that happens not to be completely ready at any given time, I will, upon
request, make every effort to complete the documentation of that field,
usually within a matter of hours.  If for whatever reason an interested
photometrist desires standard stars selected on the basis of criteria
other than those that I have used, I will do my best to provide a
customized standard list.

\vspace{5 mm}

I am very grateful to the Canadian Astronomy Data Centre and the Isaac
Newton Group Archive/UK Astronomy Data Centre  for the many valuable and
extensive public-domain data sets they have provided me.  I would like
also to thank the many individuals who have freely contributed their
proprietary data to this effort, including most particularly
Peter~Bergbusch, Mike~Bolte, Howard~Bond, Pat~Dowler, Mike~Pierce,
Alfredo~Rosenberg, Nancy~Silbermann, and Nick~Suntzeff, plus anyone else
whose name I have momentarily forgotten to mention.  We are all much
indebted to Arlo~Landolt for his many years of strenuous and punctilious
effort on our behalf.

\clearpage

\centerline{\bf FIGURE CAPTIONS}

\figcaption[fig1.ps] {The differences between my photometry in the $B$ bandpass and that of
Landolt (1992) versus magnitude (upper panel) and color (lower panel).  In each panel the
abscissa is the average of my and Landolt's photometry, and the ordinate is in the sense
my magnitude {\it minus\/} Landolt's.\label{fig1}}

\figcaption[fig1.ps] {The differences between my photometry in the $V$ bandpass and that of
Landolt (1992) versus magnitude (upper panel) and color (lower panel).  In each panel the
abscissa is the average of my and Landolt's photometry, and the ordinate is in the sense
my magnitude {\it minus\/} Landolt's.\label{fig2}}

\figcaption[fig1.ps] {The differences between my photometry in the $R$ bandpass and that of
Landolt (1992) versus magnitude (upper panel) and color (lower panel).  In each panel the
abscissa is the average of my and Landolt's photometry, and the ordinate is in the sense
my magnitude {\it minus\/} Landolt's.\label{fig3}}

\figcaption[fig1.ps] {The differences between my photometry in the $I$ bandpass and that of
Landolt (1992) versus magnitude (upper panel) and color (lower panel).  In each panel the
abscissa is the average of my and Landolt's photometry, and the ordinate is in the sense
my magnitude {\it minus\/} Landolt's.\label{fig4}}

\clearpage

\begin{table}
\begin{center}
\caption{Mean Magnitude Differences, Present {\it minus\/} Previous Photometry}
\vspace{5 mm}
\begin{tabular}{lcccc}\hline\hline
             & $\Delta B$ & $\Delta V$ & $\Delta R$ & $\Delta I$ \\
Source       & $\sigma$   & $\sigma$   & $\sigma$   & $\sigma$ \\
             & $N$        & $N$        & $N$        & $N$ \\
\hline
Landolt & --0.0005 & +0.0002 & --0.0010 & --0.0002 \\
        & \phantom{--}0.0010 & \phantom{+}0.0006 & \phantom{--}0.0011 & \phantom{--}0.0012 \\
        & 326 & 324 & 154 & 225 \\
\hline
Graham & --0.0026 & +0.0016 & --0.0128 & --0.0049 \\
             & \phantom{--}0.0019 & \phantom{+}0.0017 & \phantom{--}0.0036 & \phantom{--}0.0038 \\
             & 71         & 70         & 12          & 31 \\
\hline
Harris & --0.0056 & --0.0064 & \ldots & \ldots \\
                & \phantom{--}0.0034   & \phantom{--}0.0031   & \ldots & \ldots \\
                & 80       & 80       & \ldots & \ldots \\
\hline
Davis & --0.0015 & +0.0020  & --0.0066 & +0.0096 \\
(except \ngc{7790}) & \phantom{--}0.0025 & \phantom{+}0.0016 & \phantom{--}0.0021 & \phantom{+}0.0021\\
                & 97       & 97       & 96       & 97 \\
\hline
total stars     & 574 & 571 & 262 & 353 \\
\hline
\end{tabular}
\end{center}
\end{table}

{\renewcommand{\arraystretch}{0.78}
\begin{table}
\begin{center}
\caption{Selected Standard Fields}
\begin{tabular}{lcccrrrrl}\hline\hline
Field name   & $\alpha_{2000}$ & $\delta_{2000}$ & field~size & $N_B$ & $N_V$ & $N_R$ & $N_I$ & field type \\
\hline
T Phe        & 00 30 23.2 & --46 29 00 & 9\Min1 $\times$ 8\Min8 &    22 &   26 &    5 &   23 & Landolt field \\
SA 92        & 00 54 41.3 &  +00 40 38 & 12.6 $\times$  8.8 &    14 &   10 &    7 &   15 & Landolt field \\
PG0231+051   & 02 33 45.2 &  +05 17 51 &  8.3 $\times$  5.2 &    19 &   10 &    7 &   19 & Landolt field \\
SA 95        & 03 53 21.0 & --00 01 10 & 14.3 $\times$ 13.1 &    38 &   18 &   10 &   41 & Landolt field \\
E3           & 06 42 50.6 & --45 09 10 & 12.4 $\times$ 12.5 &    31 &   24 &   28 &   13 & Graham field \\
SA 98        & 06 51 58.2 & --00 21 05 & 18.4 $\times$ 15.3 &   386 &  276 &  358 &  281 & Landolt field \\
Rubin 149    & 07 24 16.0 & --00 32 53 &  6.9 $\times$  7.3 &   105 &  114 &    7 &  100 & Landolt field \\
\ngc{2419}   & 07 38 05.9 &  +38 54 00 & 14.9 $\times$ 12.5 &    84 &  771 &  136 &  767 & globular cluster \\
\ngc{2437}   & 07 41 47.6 & --14 48 54 & 14.3 $\times$ 14.3 &   402 &  143 &  377 &   75 & open cluster \\
\ngc{2818}   & 09 16 05.6 & --36 36 39 & 14.8 $\times$ 14.6 &   710 &  178 &  709 &    0 & open cluster \\
E4           & 09 23 41.8 & --45 25 11 & 11.6 $\times$  8.8 &   144 &  115 &  140 &    4 & Graham field \\
SA 101       & 09 57 29.8 & --00 21 16 & 11.6 $\times$ 11.2 &    43 &    4 &   44 &    9 & Landolt field \\
E5           & 12 05 06.0 & --45 33 18 & 12.5 $\times$  9.1 &   140 &  140 &    2 &    7 & Graham field \\
\ngc{4147}   & 12 10 12.2 &  +18 31 24 &  9.7 $\times$  9.3 &   135 &  205 &  200 &  175 & globular cluster \\
PG1323--086  & 13 25 50.8 & --08 50 04 &  7.4 $\times$  6.7 &    16 &   25 &    4 &   24 & Landolt field \\
\ngc{5194}   & 13 28 05.2 &  +46 51 18 & 24.4 $\times$ 30.7 &    26 &   27 &   21 &    6 & spiral galaxy \\
\ngc{5272}   & 13 41 33.3 &  +28 22 10 &  5.4 $\times$  5.0 &   189 &  241 &    0 &  226 & globular cluster \\
\ngc{5904}   & 15 18 21.2 &  +02 04 53 & 17.6 $\times$ 11.0 &   627 &  535 &    0 &  643 & globular cluster \\
PG1633+099   & 16 35 32.1 &  +09 47 55 &  4.0 $\times$  3.5 &    32 &   47 &    5 &   47 & Landolt field \\
PG1657+078   & 16 59 35.7 &  +07 42 30 &  5.6 $\times$  5.4 &    32 &   51 &   44 &   51 & Landolt field \\
\ngc{6341}   & 17 17 06.5 &  +43 07 26 & 11.6 $\times$ 13.6 &  3119 & 2024 & 1477 & 3127 & globular cluster \\
Draco        & 17 19 46.3 &  +57 55 26 & 10.1 $\times$  5.8 &   190 &  168 &  160 &  116 & dwarf galaxy \\
E7           & 17 27 19.6 & --45 01 47 &  5.0 $\times$  2.9 &   121 &  121 &    2 &    2 & Graham field \\
\ngc{6633}   & 18 27 21.2 &  +06 33 19 & 28.1 $\times$ 30.5 &   180 &  163 &    0 &  198 & open cluster \\
SA 110       & 18 40 48.5 &  +00 01 51 &  7.1 $\times$  6.5 &    11 &   50 &    7 &   50 & Landolt field \\
E8           & 20 07 28.8 & --44 42 03 &  6.7 $\times$  4.6 &    33 &   33 &    2 &    2 & Graham field \\
Markarian A  & 20 43 59.0 & --10 47 01 &  5.0 $\times$  5.1 &    22 &   30 &    2 &   30 & Landolt field \\
\ngc{7006}   & 21 32 28.8 & --01 06 18 &  6.6 $\times$  4.4 &    59 &   59 &   20 &    0 & globular cluster \\
\ngc{7089}   & 21 33 47.6 & --00 51 06 &  7.4 $\times$  7.8 &   266 &   28 &    0 &  255 & globular cluster \\
SA 113       & 21 40 58.9 &  +00 27 43 &  5.2 $\times$  4.9 &    30 &   34 &    4 &   30 & Landolt field \\
BL Lac       & 22 02 41.0 &  +42 16 57 &  5.3 $\times$  5.5 &    53 &   53 &    0 &    0 & AGN \\
\ngc{7790}   & 23 58 26.1 &  +61 13 00 &  6.7 $\times$  6.1 &   238 &  240 &  243 &  238 & open cluster \\
\hline
\hline
\end{tabular}
\end{center}
\end{table}}
\end{document}